\documentclass[a4paper]{article}

\usepackage{INTERSPEECH2020}
\usepackage{multirow}
\usepackage{float}
\usepackage{tabularx}
\usepackage{csvsimple}
\usepackage{mathtools}
\usepackage{amssymb}
\usepackage{amsmath}
\usepackage{hyperref}
\usepackage{commath}
\usepackage{graphicx}
\usepackage{subcaption}
\usepackage{cite}

\DeclarePairedDelimiter\ceil{\lceil}{\rceil}

\begin{document}

\title{CopyCat: Many-to-Many Fine-Grained Prosody Transfer for Neural Text-to-Speech}
\name{Sri Karlapati, Alexis Moinet, Arnaud Joly, Viacheslav Klimkov, Daniel S\'{a}ez-Trigueros, Thomas~Drugman}
\address{Amazon Research, Cambridge, United Kingdom}
\email{srikarla, amoinet, jarnaud, vklimkov, dsaez, drugman@amazon.com}

\maketitle

\begin{abstract}
    Prosody Transfer (PT) is a technique that aims to use the prosody from a source audio as a reference while synthesising speech. Fine-grained PT aims at capturing prosodic aspects like rhythm, emphasis, melody, duration, and loudness, from a source audio at a very granular level and transferring them when synthesising speech in a different target speaker's voice. Current approaches for fine-grained PT suffer from source speaker leakage, where the synthesised speech has the voice identity of the source speaker as opposed to the target speaker. In order to mitigate this issue, they compromise on the quality of PT. In this paper, we propose CopyCat, a novel, many-to-many PT system that is robust to source speaker leakage, without using parallel data. We achieve this through a novel reference encoder architecture capable of capturing temporal prosodic representations which are robust to source speaker leakage. We compare CopyCat against a state-of-the-art fine-grained PT model through various subjective evaluations, where we show a relative improvement of $47\%$ in the quality of prosody transfer and $14\%$ in preserving the target speaker identity, while still maintaining the same naturalness. 
    
\end{abstract}
\noindent\textbf{Index Terms}: Neural text-to-speech, fine-grained prosody transfer, many-to-many prosody transfer.

\section{Introduction}
    In recent times, neural text-to-speech (NTTS) methods have significantly improved the naturalness of synthesised speech obtained from TTS systems\cite{sotelo2017char2wav,wang2017tacotron,gibiansky2017deep,shen2018natural,li2018close,vandenOord2016Wavenet,kalchbrenner2018WaveRNN,lorenzo2018robust}. In this paper, by NTTS systems, we refer to a subset of NTTS systems that predict mel-spectrograms from text, followed by a neural vocoder as proposed in \cite{shen2018natural}. As an extension to NTTS, neural prosody transfer techniques \cite{akuzawa2018expressive,skerry2018towards,zhang2018learning,wang2018style,lee2018robust,klimkov2019finegrained} were introduced which use the prosody from a source audio as a reference when synthesising speech, for example, in a different, target speaker's voice.

    There have been several approaches proposed for prosody transfer (PT), and they can be classified into: 1) Coarse-Grained Prosody Transfer (CPT) techniques \cite{akuzawa2018expressive,skerry2018towards,zhang2018learning,wang2018style} and 2) Fine-Grained Prosody Transfer (FPT) techniques \cite{lee2018robust,klimkov2019finegrained}. While CPT techniques focus on capturing sentence-level prosodic features like style or emotion, which can be transferred across sentences of different text, FPT techniques focus on capturing prosodic features like rhythm, emphasis, melody, and loudness, which can not necessarily be transferred between sentences of different text. Both CPT and FPT techniques get latent representations from either, a mel-spectrogram \cite{akuzawa2018expressive,skerry2018towards,zhang2018learning,wang2018style,lee2018robust} or hand-crafted features \cite{klimkov2019finegrained} known to have a strong correlation with prosody, and use them to condition the NTTS system. In CPT methods, the latent representation is in the form of a single time-independent vector, while in FPT methods, time-dependent latent representations are obtained. The time-dependency of latent representations in FPT can be either at the phoneme level or the mel-spectrogram frame-level \cite{lee2018robust}. In our work, we propose a novel, frame-level FPT model, capable of transferring prosody from any source speaker to a fixed set of target speakers.

    While training, PT methods generally use latent representations and speaker embeddings obtained from a reference mel-spectrogram, to condition the NTTS system predicting the same mel-spectrogram. At inference, the speaker embedding of a different speaker is used to condition the NTTS system, to synthesise speech in a target speaker's voice. Since the input to getting the conditioning and the output from NTTS is the same whilst training, if the latent representations have enough capacity, the NTTS system can learn to depend on the latent representations for deciding the speaker identity with which it should generate the output and can ignore the speaker embedding. During inference, this results in the synthesised speech having more of the source speaker's identity than the target speaker's identity; we refer to this phenomenon as source speaker identity leakage. While CPT methods have lesser capacity owing to the time-independent nature of their latent representation, they have limitations in the granularity of PT they can achieve. FPT methods tend to have more capacity and they can achieve PT of very fine granularity; however, they have an increased chance of source speaker leakage. Our work proposes a novel NTTS architecture for FPT that is robust to source speaker leakage while still obtaining a high quality of fine-grained PT.
    
    In this work, we propose CopyCat (CC), a novel, fully parallel, frame-level, FPT model capable of transferring prosody from any source speaker to a fixed set of target speakers, while being robust to source speaker leakage. We make the model fully-parallel by using oracle phoneme durations obtained through forced alignment between the text and source audio, as opposed to using an attention mechanism commonly used in sequence to sequence models. We compare our method to \cite{klimkov2019finegrained} through various evaluations. We show that CC obtains a relative improvement of 47\% in the quality of PT, and 14\% in maintaining the target speaker's identity, while preserving the same level of naturalness. Our subjective evaluations show that para-linguistics like breaths are being transferred by CC which is not possible in \cite{klimkov2019finegrained}.

\section{CopyCat}
    \begin{figure*}
        \centering
        \begin{subfigure}{0.34\linewidth}
            \includegraphics[width=\linewidth]{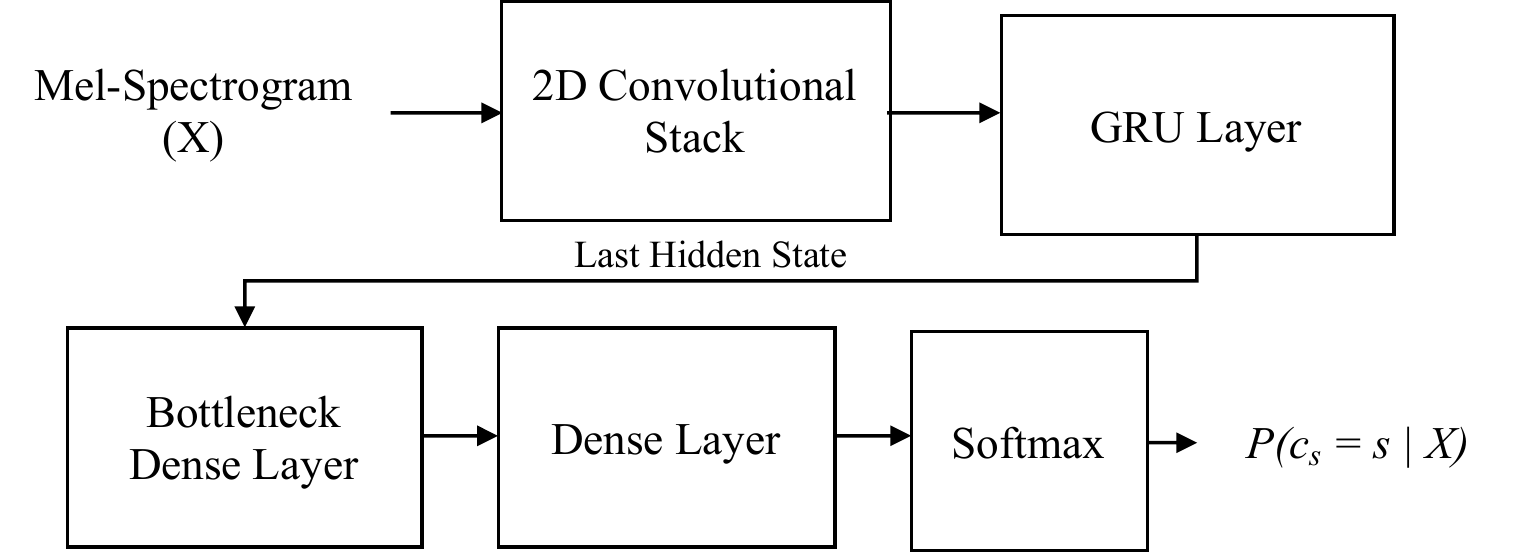}
            \caption{Speaker Classifier Architecture}
            \label{fig:speaker_classifier}
        \end{subfigure}
        \hspace*{\fill} 
        \begin{subfigure}{0.65\linewidth}
            \includegraphics[width=\linewidth]{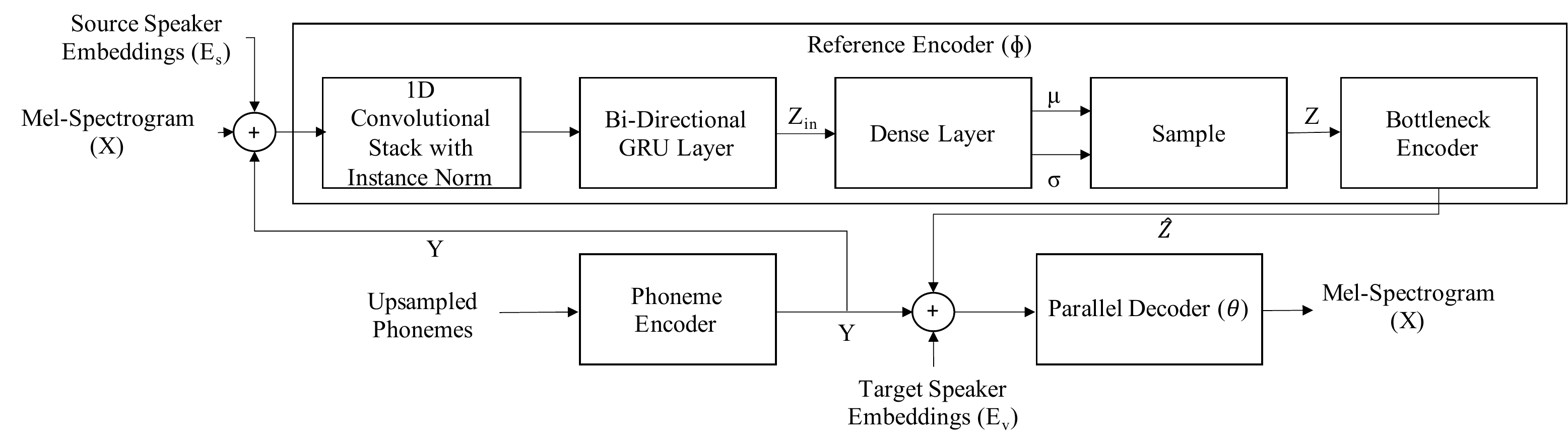}
            \caption{CopyCat Model Architecture}
            \label{fig:model_architecture}
        \end{subfigure}
        \vspace{-0.3cm}
        \caption{Figure~(a) shows the architecture of the speaker classifier used to obtain the speaker embeddings used by the CopyCat model. Figure~(b) shows the architecture of the proposed CopyCat model and its three main components, namely, the reference encoder ($\phi$), the phoneme encoder, and the parallel decoder ($\theta$).}
        \vspace{-0.5cm}
    \end{figure*}
    Our CopyCat model takes upsampled phonemes, the corresponding mel-spectrogram, and speaker embeddings, as input. In Section~\ref{sec:speaker_classifier}, we describe the architecture of the speaker classifier used to get the speaker embeddings. In Section~\ref{sec:model_architecture}, we describe the architecture of the CopyCat model and how it uses the aforementioned inputs. We describe the training and inference methodology in Section~\ref{sec:training_inf}.
 
    \subsection{Speaker Classifier}
        \label{sec:speaker_classifier}
        CopyCat uses speaker embeddings for identifying the speaker identity with which it must generate the output. To obtain these embeddings, we trained a speaker classifier with a bottleneck as shown in Figure~\ref{fig:speaker_classifier}. The model takes a mel-spectrogram as input and provides logits for each speaker class as output. The time-independent representation it learns at the bottleneck layer is used as the speaker embedding, $E_s$, for a given mel-spectrogram, $X$. We use a stack of 2D convolutional layers to reduce the dimensions along both the time and frequency axes, followed by a GRU layer along the shortened time axis. We use the last state from the GRU layer and pass it through the bottleneck dense layer and use the resultant output as, $E_s$. We project $E_s$ to match the number of speakers in $\{ S\}$, using a dense layer and apply softmax to get the probability, $a_s=p(c_s=s\mid X) \ \forall s\in S$, where $c_s$ is the speaker identity of $X$. We use cross-entropy loss \cite{murphy2012MLBook} to train the model.
    
    \subsection{Model Architecture}
        \label{sec:model_architecture}
        The CC model in Figure~\ref{fig:model_architecture} consists of three components, 1) Phoneme Encoder that learns phoneme encodings, 2) Reference Encoder that learns prosodic representations, and 3) Parallel Decoder that generates mel-spectrograms using the phoneme encodings, prosodic representations, and speaker embeddings.
        \subsubsection{Phoneme Encoder}
            \label{sec:phoneme_encoder}
            The phoneme encoder has a similar architecture to the character encoder in \cite{shen2018natural}. The input phonemes are upsampled using forced alignment to match $T$, the number of frames in $X$. We do this by aligning the phonemes and audio as described in \cite{klimkov2019finegrained}, which provides us with the duration of each phoneme. Using these durations, we upsample the phonemes to match the number of mel-spectrogram frames aligned with them. These upsampled phonemes are given as input to this layer to get phoneme encodings, $Y=[\vec{y}^0, \vec{y}^1, \dots, \vec{y}^{T-1}]$.

        \subsubsection{Reference Encoder}
            The reference encoder obtains temporal prosodic representations from a given mel-spectrogram, $X$. We use these representations to condition the parallel decoder. The reference encoder consists of the following blocks.\medskip

            \noindent\emph{Convolutional Layers with Instance Norm:} 
                Instance normalisation can be used to reduce the presence of stationary features like speaker identity from $X$ \cite{chou2019one}. It is based on the hypothesis that the constant factor in each of the channels is speaker identity, which can be removed by normalising each channel by its mean and standard deviation. Firstly, we take the mean and standard deviation along each channel axis after convolution,
                \begin{equation}
                    \mu_c = \frac{1}{N} \sum_{i=0}^{N-1} K_c[i] , \ \forall c \in C,
                    \label{eq:in_norm_mean}
                \end{equation}
                \begin{equation}
                    \sigma_c = \sqrt{\frac{1}{N} \sum_{i=0}^{N-1} (K_c[i] - \mu_c)^2} , \ \forall c \in C,
                    \label{eq:in_norm_var}
                \end{equation}
                where $K_c \in \mathbb{R}^{U\times V}$ is the output from the $c$-th channel after convolution, $C$ is the set of all output channels, and $N = U*V$, the number of elements in $K_c$. Then, we normalise each element in $K_c$,
                \begin{equation}
                    K_c'[w] = \frac{K_c[w] - \mu_c}{\sigma_c} , \ \forall c \in C.
                    \label{eq:in_norm}
                \end{equation}
                We applied 3 of such convolution layers, followed by a bi-directional GRU, whose hidden-states we represent as $Z_{in}\in \mathbb{R}^{T\times H}$, where $H$ is the size of the hidden states. We refer to the set of all parameters used in this block to be $\{\gamma\}$. 
                
                When mel-spectrograms of audio samples that sound significantly different from the training data are given, $Z_{in}$ tends to be sparse, making it difficult for the decoder to transplant the prosody on a new speaker's voice. To mitigate the sparseness, we made this block a variational encoder, as described below.\medskip

            \noindent\emph{Conditional Variational Encoding:}
                To get a dense time-dependent latent prosodic representation, $Z\in \mathbb{R}^{T\times H}$, we condition both the encoder, $q_\phi(Z\mid X, Y, E_s)$, and the decoder, $p_\theta(X\mid Y, Z, E_s)$, in a VAE \cite{kingma2013auto,sohn2015learning} with the same conditions. We assume a prior distribution, $p(\vec{z}^i) = \mathcal{N}(\vec{z}^i; 0, I)\forall\ \vec{z}^i \in Z$. We define $\{\gamma\} \subset \{\phi\}$ and train the model to maximize the evidence lower bound (ELBO) defined in Equation~\ref{eq:kld}, where $\alpha$ is used as the anneal factor to avoid posterior collapse \cite{sonderby2016train}.
                \begin{multline}
                    \mathbb{L}(p_\theta, q_\phi) = \mathbb{E}_{q_\phi(Z\mid X)}[log(p_\theta(X\mid Y, Z, E_s))] \\ - \alpha \sum_{i=0}^{T-1} D_{KL}(q_\phi(\vec{z}^i\mid X, Y, E_s) \mid \mid p(\vec{z}^i)).
                    \label{eq:kld}
                \end{multline}
            
            \noindent\emph{Bottleneck Encoder:}
                During training, the decoder can learn to depend on $Z=[\vec{z}^0, \vec{z}^1, \dots, \vec{z}^{T-1}]$ to provide the speaker identity while disregarding other conditionings. This results in source speaker identity leakage. Empirically, we found that if the capacity of the hidden dimension, $H$, was too small, it degraded the quality of PT, while a large $H$ resulted in source speaker leakage. We noted that prosody can be considered to vary across a few frames. Therefore, we introduced a temporal bottleneck \cite{qian2019autovc}, which reduces the amount of information flowing from the reference encoder to the decoder along the time axis. It forces the decoder to depend on other conditioning for the phonetic and speaker identity information while depending on the output from the bottleneck encoder just for prosodic information.
                
                We downsample $Z$ along the time axis at a fixed rate $\tau$ to get $Z^{\downarrow} \in \mathbb{R}^{\ceil*{T/\tau} \times H}$ and then upsample $Z^{\downarrow}$ to length $T$ to get $\hat{Z}\in \mathbb{R}^{T\times H}$. Since the input to this layer is $Z$, which is sampled from the hidden states of a bi-directional GRU, we first split $Z$ along the hidden dimension into matrices, $Z_{\rightarrow}$ and $Z_{\leftarrow}\ \in \mathbb{R}^{T\times H/2}$, to represent the forward and backward hidden states, respectively. We now get, $Z_{\rightarrow}^{\downarrow}=[\vec{z_{\rightarrow}^{\tau-1}}, \vec{z_{\rightarrow}^{2\tau-1}}, \dots]$, and $Z_{\leftarrow}^{\downarrow}=[\vec{z_{\leftarrow}^{0}}, \vec{z_{\leftarrow}^{\tau}}, \vec{z_{\leftarrow}^{2\tau}}, \dots]$. We concatenate $Z_{\leftarrow}^{\downarrow}$ and $Z_{\rightarrow}^{\downarrow}$ along the hidden dimension to get $Z^{\downarrow}$ which we upsample by replication at the rate $\tau$ to get $\hat{Z}$. 
            
        \subsubsection{Parallel Decoder}
            The parallel decoder consists of a stack of 3 convolutional layers followed by a bi-directional GRU. The decoder predicts the output mel-spectrogram, $X=[\vec{x}^0, \vec{x}^1, \dots, \vec{x}^{T-1}]$, given the phoneme encodings, $Y$, the latent representation, $\hat{Z}$, and the speaker embedding, $E_s$. This can be represented by modifying the first term in Equation~\ref{eq:kld} as,
            \begin{equation}
                \label{eq:parallel_decoder}
                p_{\theta}(X\mid Y,\hat{Z},E_s) = \prod_{t=0}^{T-1} p_{\theta}(\vec{x}^t\mid Y,\hat{Z},E_s).
            \end{equation}
            As both the input and output from the decoder are of the same length, there is no need for an attention layer to align the sequences. Since, we have information available both in the forward and backward directions through the bi-directional GRU, we do not introduce auto-regression to avoid biasing the model in a particular direction of decoding.

        \subsection{Discriminator}
            To improve the segmental quality of the samples produced by our CC model, and further maximize the ELBO, we introduce a discriminator $D$ to fine-tune the model using adversarial training. We use the self-attention discriminator proposed in \cite{zhang2019self} and the hinge version of the adversarial loss \cite{lim2017geometric}, both of which have provided good results in image generation \cite{brock2018large} and text-to-speech \cite{guo2019new} tasks, as shown below.
            \begin{equation}
                \begin{split}
                    \mathbb{L}_D = & - \mathbb{E}_{X \sim p_{data}} [\min(0, -1+D(X))] \\
                                   & - \mathbb{E}_{\hat{X} \sim p_{\theta}} [\min(0, -1-D(\hat{X}))]. \\
                    \mathbb{L}_G = & -\mathbb{E}_{\hat{X} \sim p_{\theta}} [D(\hat{X})].
                    \label{eq:generator_loss}
                \end{split}
            \end{equation}
            Here, $G = \{\phi\} \cup \{\theta\}$ is the generator. It is composed of the encoder $\phi$ and the decoder $\theta$ defined in Section~\ref{sec:model_architecture}. $X$ are real mel-spectrograms and $\hat{X}$ are synthetic mel-spectrograms produced by $G$. Since our goal is to improve segmental quality, we encourage the discriminator to pay attention to short-term transitions in the audio by feeding a random window of 32 mel-spectrogram frames as input. Owing to the small random window size, this approach has the advantage of being faster and robust to overfitting when compared to methods using full mel-spectrograms, as the discriminator can then fed with a larger input batches.

    \subsection{Training \& Inference Methodology}
        \label{sec:training_inf}
        \subsubsection{Training}
            We train the model in 2 steps, 1) initial training and 2) adversarial fine-tuning. During both the steps, the reference mel-spectrogram and the output mel-spectrogram are the same. During the initial training step, we train the model by replacing the first term in Equation~\ref{eq:kld} by Equation~\ref{eq:parallel_decoder}. We linearly increase the $\alpha$ coefficient of the KL-divergence term from $0$ to $1$. During the adversarial fine-tuning step, we add the generator loss term, $\mathbb{L}_G$, in Equation~\ref{eq:generator_loss} to the loss from the initial training stage and the discriminator is trained using, $\mathbb{L}_D$, in Equation~\ref{eq:generator_loss}.

        \subsubsection{Inference}
            During inference, instead of using the same $E_s$ as input to the encoder and the decoder, as done in training, we condition the decoder to use the centroid of the target speaker, $E_{v_c}=\mid V\mid^{-1} \sum_{v\in V}E_v$, where $V$ is the set of all utterances in the training set of a target speaker. Note that the input to the encoder remains unchanged. Since we trained a speaker classifier, the embeddings for mel-spectrograms with similar speaker identities are placed close to each other in the embedding space and so the centroid is chosen to maintain a consistent speaker identity across synthesised sentences in the target speaker's voice.

\section{Experiments and results}
    \subsection{Data}
        We conducted experiments on an internal US English dataset of long-form recordings. The training dataset consisted of a combined total of 35 hours of long-form data recorded by 5 female speakers. A combined total of 6.5 hours of long-form data from various sources recorded by 12 female speakers was used as the test dataset. Those 12 speakers were not a part of the training dataset. We also had our test dataset recorded by speaker from the training dataset which was used only for the naturalness evaluation and not to train our models.
        
        To train the speaker classifier, we used the LibriSpeech dataset \cite{panayotov2015librispeech}, with 1000 speakers. None of the speakers in our training set or test set were part of the LibriSpeech dataset.

    \subsection{Evaluations} 
        We evaluated CopyCat against a state-of-the-art model in FPT \cite{klimkov2019finegrained}. We trained both a single-speaker aggregated prosody model (SSAP) as described in \cite{klimkov2019finegrained}, and a multi-speaker aggregated prosody model (MSAP), by conditioning the decoder in SSAP with the same speaker embeddings that are used by CC. We compared these systems through 3 different evaluations and used $p-values$ from pairwise two-sided Wilcoxon signed-rank tests to evaluate the statistical significance of the results.

        \begin{figure*}
            \centering
            \begin{subfigure}{0.33\linewidth}
                \includegraphics[width=\linewidth]{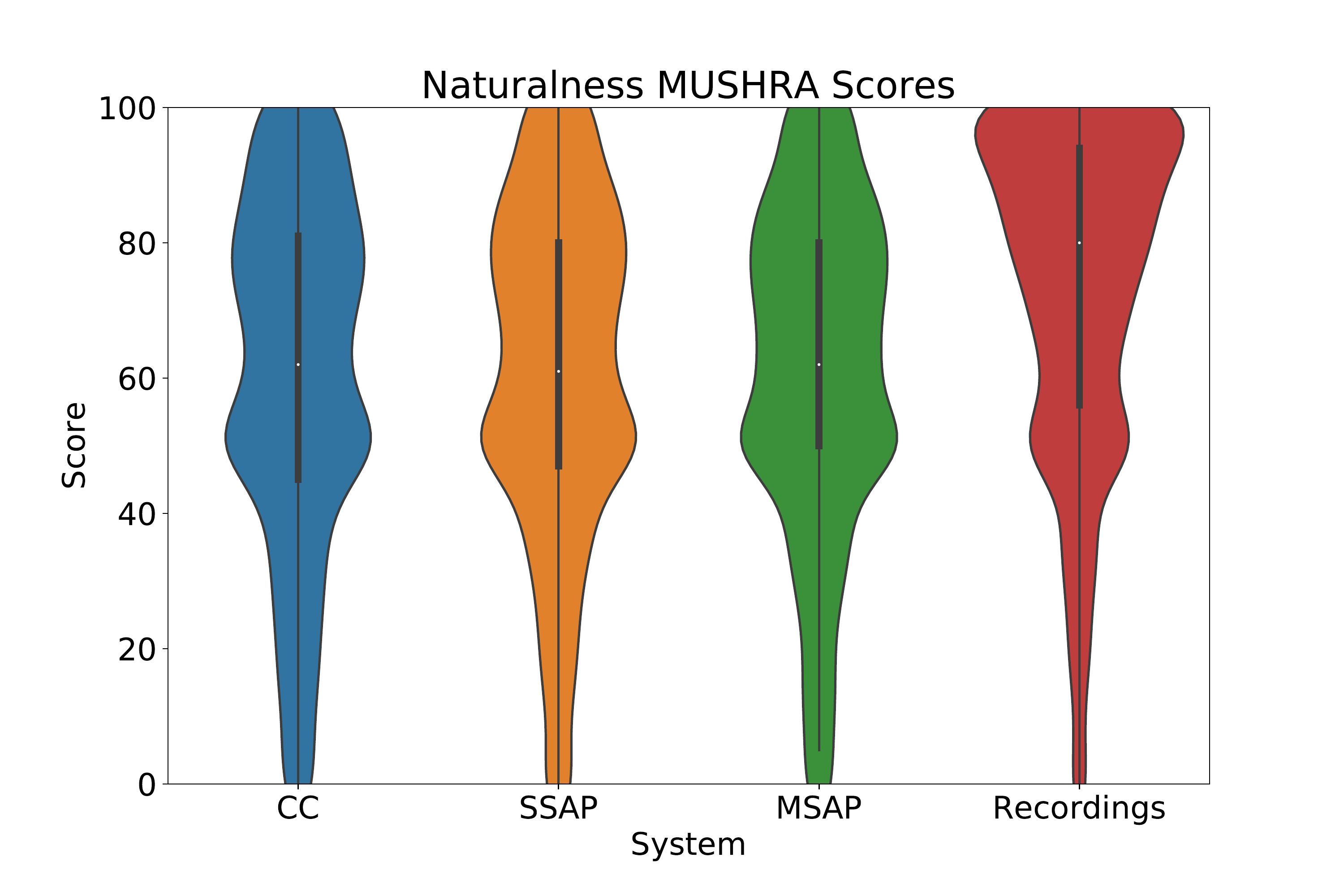}
                \vspace{-0.7cm}
                \caption{Naturalness}
                \label{fig:mushra_naturalness}
            \end{subfigure}
            \begin{subfigure}{0.33\linewidth}
                \includegraphics[width=\linewidth]{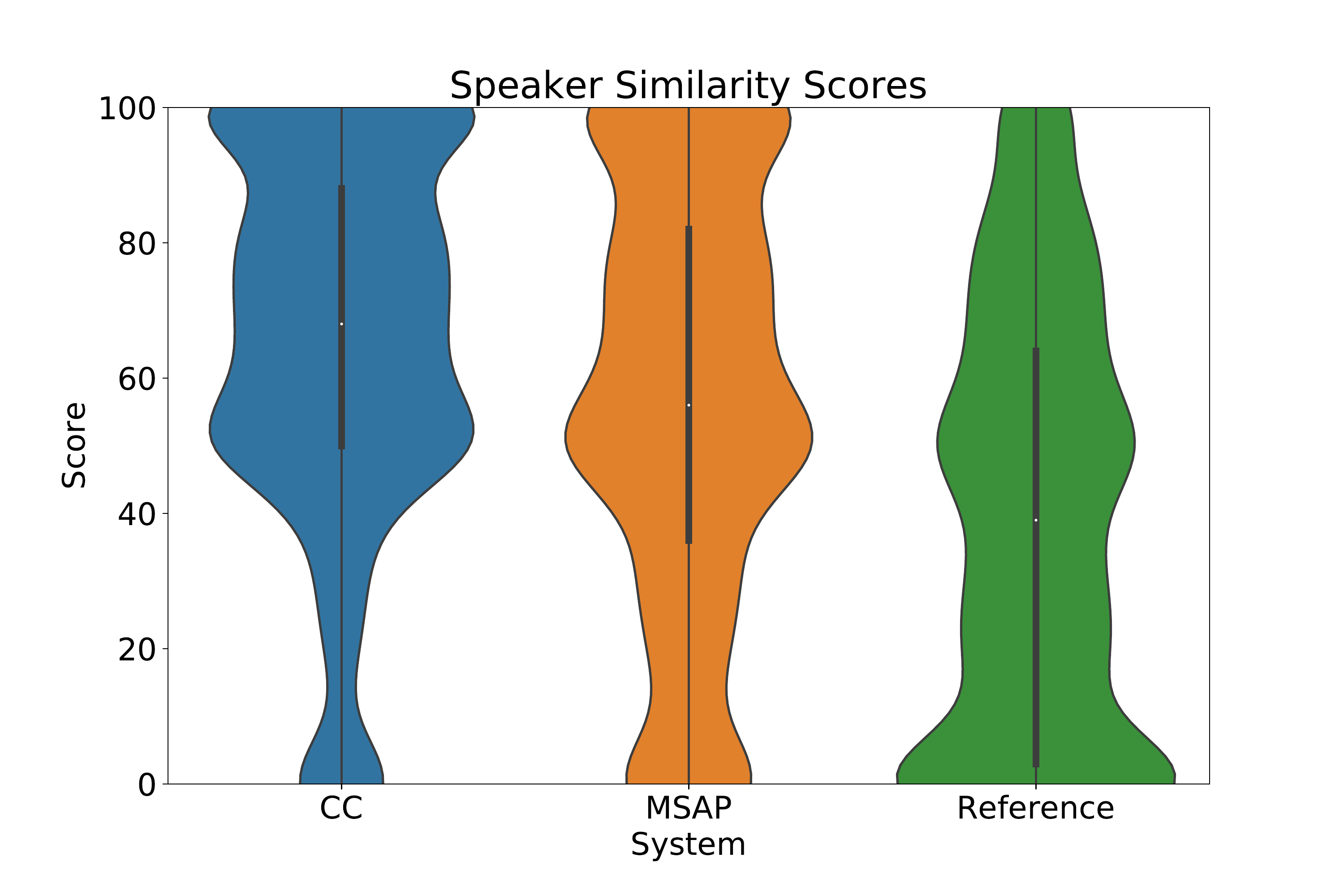}
                \vspace{-0.7cm}
                \caption{Speaker Similarity}
                \label{fig:mushra_speaker_similarity}
            \end{subfigure}
            \begin{subfigure}{0.33\linewidth}
                \includegraphics[width=\linewidth]{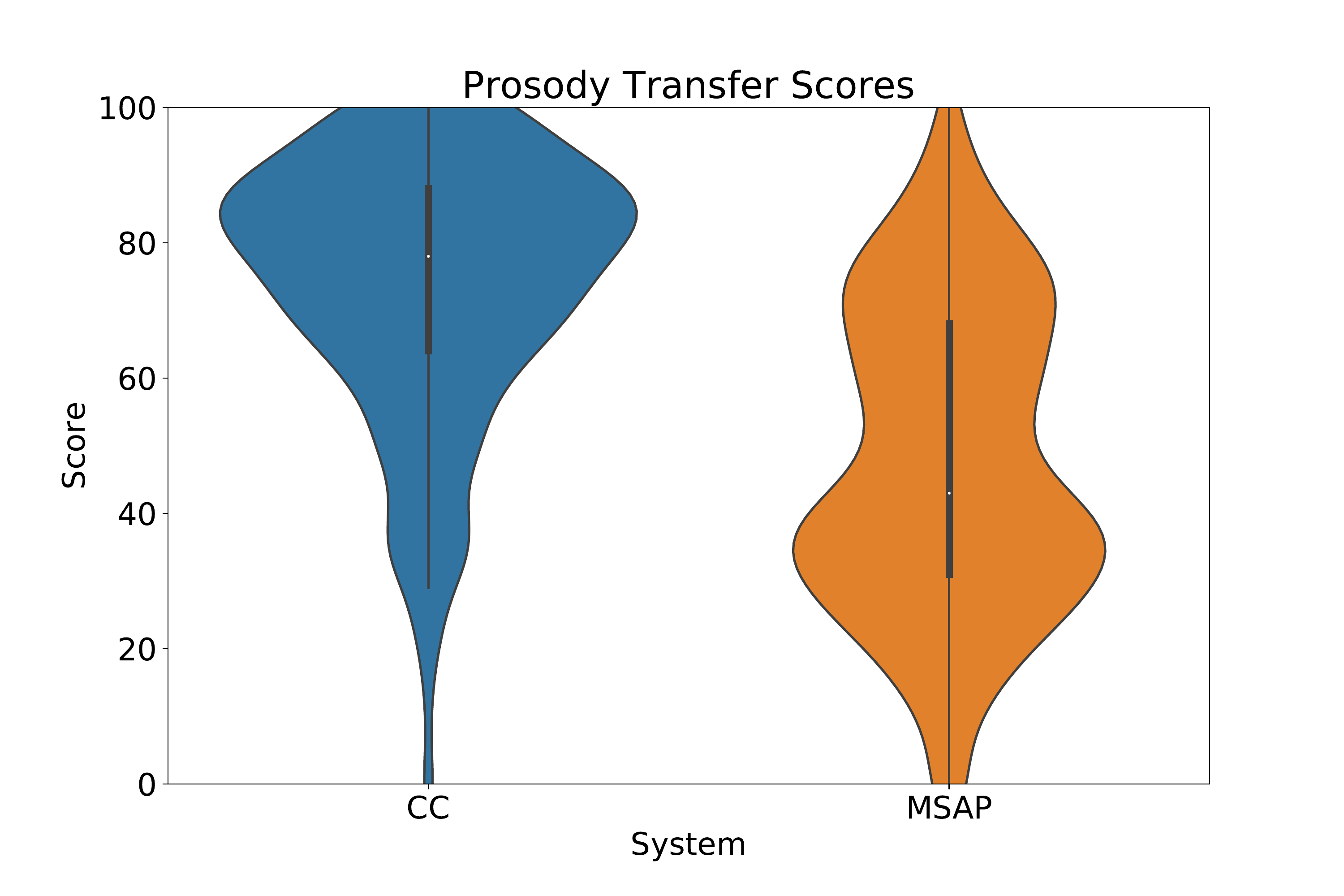}
                \vspace{-0.7cm}
                \caption{Prosody Transfer}
                \label{fig:mushra_prosody_transfer}
            \end{subfigure}
            \vspace{-0.2cm}
            \caption{Violin plots of scores obtained in the evaluations}
            \vspace{-0.5cm}
        \end{figure*}
        \begin{table}
            \centering
            \begin{tabular}{|l|c|c|c|c|}
                \hline
                \textbf{Evaluation Task}             & \multicolumn{1}{l|}{\textbf{CC}} & \multicolumn{1}{l|}{\textbf{MSAP}} & \multicolumn{1}{l|}{\textbf{SSAP}} & \multicolumn{1}{l|}{\textbf{Ref}} \\ \hline
                \textit{\textbf{Naturalness}}        & 60.59                            & 60.63                              & 60.63                              & 73.59                             \\ \hline
                \textit{\textbf{Speaker Similarity}*} & 65.25                            & 56.67                              & N/A                                & 38.93                             \\ \hline
                \textit{\textbf{Prosody Transfer}*}   & 72.74                            & 47.25                              & N/A                                & N/A                               \\ \hline
            \end{tabular}
            \caption{Mean scores obtained in the evaluations. \textit{*} shows tasks in which CC obtained a statistically significant improvement.}
            \label{tab:mushra}
            \vspace{-1cm}
        \end{table}
        \subsubsection{Naturalness}
            The systems are compared through a MUSHRA test for naturalness \cite{itu20031534}. 25 American English speakers were presented with the same 100 test cases. Each test case consisted of 3 audio samples, one for each of the 3 systems, and a sample recorded by the target speaker. The listeners were asked to rate each system in terms of naturalness of speech on a scale of 0 to 100. SSAP is a many-to-one prosody transfer system, and since we had test set recordings for only 1 speaker from the training set, we evaluated naturalness only for this case. As shown in Table~\ref{tab:mushra} and Figure~\ref{fig:mushra_naturalness}, CC obtains a MUSHRA score distribution that is statistically similar ($p-value > 0.05$) to the MUSHRA score distributions of MSAP and SSAP. This is of interest as CC is a fully parallel, non-autoregressive model, and is less complex than Tacotron like models, SSAP and MSAP. Since the MUSHRA score distributions for both MSAP and SSAP are statistically similar ($p-value > 0.05$), we discarded SSAP from future evaluations. This enabled us to evaluate MSAP and CC in further evaluations while having each of the 5 speakers in the training set as target speakers.
        
        \subsubsection{Speaker Similarity}
            \label{sec:speaker_similarity}
            In this evaluation, 25 American English speakers were presented with 100 test cases; 20 test cases from each of the 5 target speakers. Each test case consisted of 3 audio samples from CC, MSAP, and the prosodic reference audio sample from the test set. The listeners were also presented with a randomly-picked audio sample of the target speaker's voice. They were asked to rate each of the samples presented on their closeness to the voice identity of the target speaker on a scale of 0 to 100. From Table~\ref{tab:mushra} and Figure~\ref{fig:mushra_speaker_similarity}, it can be seen that CC improves on MSAP by a statistically significant 14\% ($p-value < 10^{-3}$). On inspection of samples in which MSAP obtained a low score, it was found that MSAP struggled with maintaining the identity of one target speaker for whom we had the least amount of training data. We conjecture that MSAP needs more data to maintain speaker identity than CC. We believe that CC outperforms MSAP due to the parallel decoder having a simpler architecture than MSAP, and therefore requires lesser data. 

        \subsubsection{Prosody Transfer}
            \label{sec:prosody_transfer}
            We presented 9 linguists with 100 test cases; 20 test cases from each of the 5 target speakers. Each test case consisted of a reference sample from an unseen source speaker, and prosody transferred samples from CC and MSAP, in the target speaker's voice. The listeners were asked to score each of the samples on a scale of 0-100 on how closely the sample follows the reference's prosody. When rating the systems, the linguists were asked to focus on: rhythm, emphasis, syllable length, melody, and loudness. They were also asked to consider breaths as a part of prosody for this evaluation. As can be seen from Table~\ref{tab:mushra} and Figure~\ref{fig:mushra_prosody_transfer}, CC shows a statistically significant 47\% improvement over MSAP in this test ($p-value < 10^{-3}$). Since we want to evaluate the performance of the models on PT, just for this evaluation, we removed 8 test cases from the 100, in which MSAP exhibited failure modes unrelated to prosody, such as, skipping phonemes and lost attention. This was favourable to MSAP in the prosody evaluation, as opposed to our system. Both the violin plot shown in Figure~\ref{fig:mushra_prosody_transfer} and the PT scores in Table~\ref{tab:mushra} were computed after the removal of the aforementioned test cases. We hypothesize that CC out-performs MSAP because the latent representations are obtained from a mel-spectrogram, which has more information than just the pitch and energy. CC also does not suffer from errors in pitch extraction, which can cause sudden changes in prosody in MSAP. We do not normalise the latent representations, either by speaker identity or per utterance, which was a hindrance for MSAP in samples where the reference audio had changes in emotion or prosody. Transfer of breaths also helped CC get a better score than MSAP, because MSAP modelled silences in place of breaths, resulting in a change of perceived emotion.

        \subsubsection{Cycle Consistency}
            We also checked for cycle consistency in CC \cite{zhu2017CycleConsistency}. We transferred prosody from a source speaker A to target speaker B, and obtained latent representations, $\hat{Z}^{A\rightarrow B}$. Then the synthesised mel-spectrogram with speaker B's identity is re-encoded through the reference encoder to get latent representations, $\hat{Z}^{B\rightarrow A}$. We defined Cycle Consistency Loss as $\abs{\hat{Z}^{A\rightarrow B} - \hat{Z}^{B\rightarrow A}}$. The cycle consistency loss for CC over the training set was $10^{-6}\pm 6 * 10^{-7}$. This shows that the latent representations for the same linguistic content and prosody are the same irrespective of the speaker identity. While a low cycle consistency loss is a necessary, yet, insufficient condition to claim that CC disentangled source speaker identity while retaining prosody, it is a strong metric when juxtaposed with the results from previous sections to show that the representations are robust to speaker leakage.
            
\section{Conclusions}
    We proposed CopyCat, a novel many-to-many FPT method which is robust to source speaker leakage. We presented a reference encoder capable of obtaining, temporal speaker-independent prosodic representations from a mel-spectrogram. We used these prosodic representations, upsampled phoneme encodings, and speaker embeddings to condition the parallel decoder. The model was fine-tuned using a GAN-based discriminator to improve the segmental quality. CopyCat was evaluated against an existing state-of-the-art technique in prosody transfer through various evaluations, where it shows a significant improvement in the quality of prosody transfer and speaker similarity while maintaining the same level of naturalness.

    \newpage

\bibliographystyle{IEEEtran}
\bibliography{references}

\end{document}